\documentstyle[12pt]{article}

\begin{document}
\begin{flushright}
{BI-TP 93/54}
\end{flushright}
\begin{flushright}
{BGUPHYS-93/10-2}
\end{flushright}
\begin{flushright}
{October 1993}
\end{flushright}
\begin{center}
{\bf ON THE HEAT KERNEL IN COVARIANT BACKGROUND GAUGE}
\end{center}
\medskip
\begin{center}
{\bf E.I.Guendelman$^{a}$, A.Leonidov\footnote{Alexander von Humboldt Fellow}
$^{(b,c)}$, V.Nechitailo$^{(c)}$ and D.A.Owen$^{(a)}$}
\end{center}
\begin{center}
{\it (a) Physics Department, Ben Gurion University\\Beer Sheva, Israel}\\
{\it (b) Physics Department, University of Bielefeld \\Bielefeld,
Germany}\\
{\it (c) Theoretical Physics Department, P.N.Lebedev Physics
Institute\\Moscow, Russia}\\
\medskip
\end{center}
\begin{abstract}
The first three coefficients in an expansion of the heat kernel of a
 nonminimal nonabelian kinetic operator taken in an arbitrary background
gauge in arbitrary space-time dimension are calculated.
\end{abstract}

\newpage

\section{Introduction}
In this letter we present an explicit calculation of the first three
coefficients in the quasiclassical expansion of the heat kernel of the
gluon kinetic operator taken in the arbitrary covariant background
field gauge. The importance of the heat kernel can hardly
be underestimated (see, e.g, the recent review [1]). In quantum field
theory the heat kernel is an extremely convenient technical tool for
computing the Green functions of the particles propagating in the
external fields and of the quantum corrections to the classical action
of the theory. In this way one can compute the contributions to the
charge renormalization and anomalies (see, e.g., [2]). The
attractiveness of the heat kernel method is also due to its important
role in the differential geometry, where the heat kernel is an
exponentiated covariant Laplacian on some manifold. The
expansion of this exponential gives the invariants of the considered
manifold. In particle physics the Laplacian on the manifold is nothing
else but a kinetic oprator of the particle propagating on it.
Unfortunately, in the general case one encounters more complicated
operators - for example, the gluon kinetic operator in a
general covariant background gauge reduces to a covariant Laplacian only in the
Feynman gauge - this being the reason for its convenience in the
calculations in nonabelian external fields. In a general covariant gauge one
deals with the so-called nonminimal operators. The computation of the
corresponding heat kernel becomes much more involved, and only a very
limited information on its expansion coefficients is available. The
first two coefficients for the case of gravity coupled to an abelian
field was considered in [3], the trace of the fourth coefficient for
Yang-Mills theory in four dimensions was computed in [4], and that in
the high derivative gravity in four dimensions in [5].

Below we propose a method allowing an explicit calculation of the heat
kernel expansion coefficients for the non-minimal operators in an
arbitrary dimensional space-time and compute the first three Seeley
coefficients for the nonabelian kinetic operator taken in an arbitrary
backrond gauge. We shall see, that the dependence of the Seeley
coefficients on the quantum gauge fixing parameter and the space-time
dimension is strongly interrelated.

The plan of the paper is the following. In the first section we
introduce the necessary notations and describe the calculational method
. In the second one we present the results for the heat kernel expansion
coefficients and comment on their relevance to the effective action
calculation. We conclude by summarizing the results.

\section{Calculational method}

We begin with introducing the basic notations. Let us
consider a second order elliptic operator $\cal{W}$ on the
$2\omega$-dimensional manifold $\cal{M}$. By definition the heat kernel
operator corresponding to $\cal{W}$ is obtained by its exponentiaton:
\begin{equation}
K(s)=exp(-{\cal{W}}s)
\end{equation}
We shall be interested in the expansion of the trace of the matrix
elements of the heat kernel taken at one space-time point.
For the second order operator this expansion has a form (see, e.g.,
[2]):
\begin{equation}
Tr<x|K(s)|x>=\sum_{k=0}^{\infty}b({\cal{W}},\omega|x)s^{-\omega+k}
\end{equation}
where the coefficients $b_{-\omega+k}$ are the so-called Seeley
coefficients. These coefficients are the invariants of the manifold
$\cal{M}$ and the trace of the heat kernel can be considered as a
generating function for these invariants. In the often considered
standard case the operator $\cal{W}$ is a covariant Laplacian on $\cal{M}$.
Let us mention that in this case the coefficients $b_{-\omega+k}$
depend on $\omega $ only trivially. We shall consider a more complicated case
and analyse the heat kernel
expansion for the kinetic operator of Yang-Mills particles taken in an
arbitrary
covariant background gauge

\begin{equation}
{\cal{W}}^{ab}_{\mu\nu}=-D^{2}(A)^{ab}\delta_{\mu\nu}-2f^{acb}G^{c}_{\mu\nu}-
(\frac{1}{\alpha}-1)D^{ac}_{\mu}D^{cb}_{\nu}
\end{equation}
where $f^{abc}$ are the structure constants of a corresponding Lie
algebra, $D_{\mu}(A)$ is a covariant
derivative containing the external field potential $A_{\mu}$, and
$G_{\mu\nu}$ is a corresponding field
strength. The coefficients of the heat kernel expansion
\begin{equation}
Tr<x|e^{-{\cal{W}}_{\mu \nu}^{ab}s}|x>=\sum_{k=0}^{\infty}b_{-\omega +k}(G_{\mu
\nu},\omega,\alpha|x) s^{-\omega +k}
\end{equation}
are gauge
invariant with respect to the transformations of the external field
potentials $A_{\mu}^{a}$. Our main interest is to trace their dependence
on the space-time dimension ${2\omega}$ and the quantum gauge fixing
parameter $\alpha$.
To calculate the functional trace in (4) we shall use the basis of plane
waves (see, e.g., [7,8]):
  $$Tr\ <x|e^{-{\cal{W}}s}|x> = Tr  \int \frac{d^{2\omega}p}
 {(2\pi)^{2\omega}}\ e^{-ipx}
<x|e^{-{\cal{W}}s}|x>e^{ipx}$$
where the trace is performed over the Lorentz and colour
indices. The $exp[ipx]$ should be pushed through the operator to the
left then
cancelling $exp[-ipx]$, all differentiation operators
in $\cal{W
}$ becoming shifted :
$\partial_{\mu} \rightarrow \partial_{\mu} + ip_{\mu}$.
Thus we have
\begin{equation}
  Tr\ e^{-{\cal{W}}s} = Tr \int \frac{d^{d}p}{(2\pi)^{d}}\
e^{-s{\cal{W}}( \partial_{\mu} \rightarrow \partial_{\mu} + ip_{\mu} )}1
\end{equation}
where the operator in the right hand side of (5) acts on 1.
  In the considered case \\
 ${\cal{W}}( \partial_{\mu} \rightarrow \partial_{\mu} + ip_{\mu})
= {\cal{W}}_{0}-i{\cal{W}}_{1}-{\cal{W}}_{2}$, where we have introduced the
following
notation:
\begin{eqnarray}
&&  {\cal{W}}_{0}^{\mu\nu} = p^{2}(P^{\mu\nu}_{\perp}+
\frac{1}{\alpha}P^{\mu\nu}_{\parallel}),\ \
P^{\mu\nu}_{\perp}=\delta^{\mu\nu}-\frac{p^{\mu}p^{\nu}}{p^{2}},\ \
P^{\mu\nu}_{\parallel} = \frac{p^{\mu}p^{\nu}}{p^{2}}, \nonumber \\
&& {\cal{W}}_{1\ \mu\nu} = 2p_{\alpha}D_{\alpha}\delta_{\mu\nu} +
\beta (p_{\mu}D_{\nu} + p_{\nu}D_{\mu}), \nonumber \\
&&{\cal{W}}_{2\ \mu\nu} = D^{2}\delta^{\mu\nu} + 2G_{\mu\nu} +
\beta D_{\mu}D_{\nu},\ \ \beta \equiv \frac{1}{\alpha}-1
\end{eqnarray}
and the colour indices can be trivially restored, for example
$G_{\mu\nu} \rightarrow G^{ab}_{\mu\nu}=f^{acb}G^{c}_{\mu\nu}$.

	To obtain the expansion of $Tre^{-(-{\cal{W}}_{0}
+i{\cal{W}}_{1}+{\cal{W}}_{2})s}$
in $s$ we use an ordinary perturbation theory:
$$
Tr e^{(-{\cal{W}}_{0}+i{\cal{W}}_{1}+{\cal{W}}_{2})s}
 = Tr K_{0}(s) +
\int_{0}^{s} ds_{1}Tr(K_{0}(s-s_{1})(i{\cal{W}}_{1}+{\cal{W}}_{2})K_{0}(s_{1}))
+
$$
\begin{equation}
\int^{s}_{0}ds_{1} \int_{0}^{s_{1}}ds_{2} Tr( K_{0}(s-s_{1})
(i{\cal{W}}_{1}+{\cal{W}}_{2})K_{0}(s_{1}-s_{2})(i{\cal{W}}_{1}+{\cal{W}}_{2})
K_{0}(s_{2})
 + \ldots
\end{equation}

where $K_{0}^{\mu\nu}(s)$ is a free propagator
\begin{equation}
K_{0}^{\mu\nu}(s)=e^{-{\cal{W}}_{0}^{\mu\nu}s} =e^{-sp^{2}}(P^{\mu\nu}_{\perp}
+e^{-s\beta p^{2}}
P^{\mu\nu}_{\parallel})
\end{equation}
The expressions for the Seeley coefficients are obtained by collecting
the terms of a given order in covariant derivatives.

\section{Seeley coefficients}

In this ection we shall compute the first three Seeley coefficients for
the operator (3) acting on a $2\omega$-dimensional space-time manifold
$R^{2\omega}$. The first coefficient $b_{-\omega}$ is read off from the zeroth
order term in the covariant derivatives in (7). We have
\begin{eqnarray}
Tr \int \frac{d^{2\omega}p}{(2\pi)^{2\omega}}(e^{-sp^{2}}P^{\mu\nu}_{\perp}+
e^{-\frac{s}{\alpha}p^{2}}P^{\mu\nu}_{\parallel}) = \nonumber \\
\frac{1}{2^{2\omega}\pi^{\omega}}\frac{N^{2}_{c}-1}{t^{\omega}}
(2\omega-[1-\alpha^{\omega}]),
\end{eqnarray}
where $ N_c $ is number of
colours thus getting for the first Seeley coefficient
\begin{equation}
b_{-\omega}=\frac{1}{2^{2\omega}\pi^{\omega}}(N^{2}_{c}-1)
(2\omega-[1-\alpha^{\omega}])
\end{equation}

Collecting the terms proportional to $D_{\mu}^{2}$, we obtain for the
second Seeley coefficient
\begin{equation}
b_{-\omega+1}=\frac{1}{2^{2\omega}\pi^{\omega}}
(2\omega-[1-\alpha^{\omega-1}])
 \{{\Gamma(\omega)-\frac{1}{\omega}\Gamma(\omega+1)\over \Gamma[\omega]} \}
N_{c} A_{\mu}^{a}(x)A_{\mu}^{a}(x)=0
\end{equation}
due to a well known relation for the $\Gamma$ functions. This had to be
expected because this contribution corresponds to a gluon mass and is
therefore not gauge ivariant with respect to the external field
transformartion. Eq. (11) proves it in all dimensions and arbitrary
covariant gauge.

The calculation of the term of fourth order in the covariant derivatives
is straightforward but very tedious. The computation was performed using
the REDUCE package. The result reads
\begin{equation}
b_{-\omega+2}=\frac{1}{2^{2\omega}\pi^{\omega}}
(2-\frac{2\omega}{12}+\frac{1}{12}[1-\alpha^{\omega-2}]) N_{c}
 G_{\mu\nu}^{a}(x)G_{\mu\nu}^{a}(x),
\end{equation}
which is remarcably independent of $\alpha$ at $\omega =2$. Taking
$\alpha=1$ (Feynman gauge) we restore the corresponding expression from
[2]. The formula (12) is the main result of our paper.

Now when we have obtained the explicit expressions for the first three
Seeley coefficioents $b_{-\omega+k}, k=0,1,2$ it is instructive to
analyze their dependence on the quantum gauge fixing parameter $\alpha$.
{}From (10-13) we see, that in all the examined orders the $\alpha$- dependent
contributions
to the heat kernel expansion have a form
\begin{equation}
(1-\alpha^{\omega-k})s^{-\omega+k}=s^{-\omega+k}-({s \over \alpha})
^{-\omega-k}
\end{equation}
Let us now recall that the (unregularized) contribution to the effective
action originating from the kinetic operator has a form
\begin{equation}
W[A]={1 \over 2}Tr Log {\cal{W}}=-{1 \over 2} \int_{0}^{\infty} {ds \over
s} Tre^{-{\cal{W}}s}
\end{equation}
Thus the $\alpha$-dependent contribution to the effective action is
proportional to
\begin{equation}
\int {ds \over s}(s^{-\omega+k}-({s \over \alpha})^{-\omega+k})
\end{equation}
We see that if (15) would be the final expression, the answer would not
depend on $\alpha$ because of the invariance of (15) with respect to the
proper time rescaling. However this scaling invariance is explicitely
broken by the regularization procedure. At $\omega=2$ the charge
renormlaization contribution from (12) will contain contribution
proportional to $Log\alpha$ in the finite part (in agreement with [4]).
 This brings in a
question of determining a background Landau gauge ($\alpha=0$) in this
situation. It seems that the only way to do it is to include the
substraction of the terms proportional to $Log\alpha$ in the definition
of the renormalization procedure (the choice of a scheme). After such a
substraction one can take a limit of $\alpha \rightarrow 0$. Landau
gauge is of a conciderable interest because it corresponds to
calculation of the quantum gauge invariant effective action [9].

\section{Conclusion}
We have proposed an algebraically convenient procedure of calculating
the seeley coefficients for the nonminimal kinetic operator for the
Yang-Mills particles taken in the arbitrary covari<nt background gauge
for arbitrary space-time dimension $2\omega$ and have computed the first
three coefficients in this expanison. We observe that the answers have a
universal structure corresponding to the rescaling of the proper time in
some terms by the quantum gauge fixing parameter $\alpha$. The
consequencies for the computation of the corresponding terms in the
effective action are briefly discussed. We think that it would be very
interesting to extend this result to higher orders in the heat kernel
expansion. Because of the purely algebraical nature of the proposed
procedure we can expest this extention to be straightforward (though
very tedious). One of the interesting questions in looking at the higher
order terms is the interrelation between the quantum gauge fixing
parameter and the infrared cutoff that is often introduced to define
otherwise infrared divergent higher order contributions to the effective
action.
\section{Acknowledgements}
A.L. is grateful to Prof.\ H.Satz for kind hospitality at the University
of Bielefeld and acknowledges the partial finansial support by the
Russian Fund for Fundamental Research, Grant 93-02-3815.
\section{References}

1. G.A.Vilkovisky. {\it{Heat Kernel: Rencontre Entre Physiciens et
Mathematiciens}}, Preprint CERN-TH.6392/92;

2. V.N.Romanov, A.S.Shwarz, {\it{Theor.Math.Phys}} {\bf{39}}(1980),967;

3.V.P.Gusynin, E.V.Gorbar and V.V.Romankov, {\it{Nucl.Phys}} {\bf{B362}}
(1991),449;

4. A.O Barvinsky, G.A.Vilkovisky, {\it{Phys.Repts}} {\bf{119}} (1985),
1;

5. E.S.Fradkin, A.A.Tseytlin, {\it{Nucl.Phys}} {\bf{B201}} (1982), 469;

 6. A.S.Scwarz, {\it{Quantum Field Theory and Topology}}, Moskva, Nauka,
1989 (in Russian);

7. M.L.Goldberger, E.N.Adams, {\it{Journ.Chem.Phys}}\ {\bf{20}} (1952),
240;

8. D.I.Diakonov,V.Yu.Petrov and A.V.Yung, {\it{Sov.J.
Nucl.Phys}} {\bf{39}} (1984), 150;

9. E.S.Fradkin, A.A.Tseytlin, {\it{Nucl.Phys}} {\bf{B234}} (1984), 509;
   A.Rebhan, {\it{Nucl.Phys}} {\bf{288}} (1987), 832.

\end{document}